\documentclass[%
reprint,
superscriptaddress,
w
amssymb,amsmath,
]
{revtex4-1}

\usepackage[english]{babel} 
\usepackage [table]{xcolor}

\usepackage{graphicx, upgreek, subfigure} 

\usepackage{float}
\usepackage{textcomp}
\usepackage{graphicx}
\usepackage{dcolumn}
\usepackage{bm}
\usepackage[mathlines]{lineno}
\usepackage{scalerel,amssymb}
\usepackage{physics}
\usepackage{array}
\usepackage[normalem]{ulem}

\newcommand{\sz}{S$_{\rm z}$}
\newcommand{\sx}{S$_{\rm x}$}
\newcommand{\dfdb}{$\partial f/\partial B_0$}
\newcommand{\dfdA}{$\partial f/\partial A$}

\begin{document}

\begin{abstract}
We investigated the use of dielectric layers produced by atomic layer deposition (ALD) as an approach to strain mitigation in composite silicon/superconductor devices operating at cryogenic temperatures. We show that the addition of an ALD layer acts to reduce the strain of spins closest to silicon/superconductor interface where strain is highest. We show that appropriately biasing our devices at the hyperfine clock transition of bismuth donors in silicon, we can remove strain broadening and that the addition of ALD layers left $T_2$ (or temporal inhomogeneities) unchanged in these natural silicon devices. 

\end{abstract}

\title
{Reducing strain in heterogeneous quantum devices using atomic layer deposition}

\author{Oscar~W.~Kennedy}
\altaffiliation{oscar.kennedy@ucl.ac.uk} 
\affiliation{London Centre for Nanotechnology, UCL, 17-19 Gordon Street, London, WC1H 0AH, UK}

\author{James~O'Sullivan}
\affiliation{London Centre for Nanotechnology, UCL, 17-19 Gordon Street, London, WC1H 0AH, UK}

\author{Christoph~W.~Zollitsch}
\affiliation{London Centre for Nanotechnology, UCL, 17-19 Gordon Street, London, WC1H 0AH, UK}

\author{Christopher~N.~Thomas}
\affiliation{Cavendish Laboratory, University of Cambridge, JJ Thomson Ave,  Cambridge CB3 0HE, UK}


\author{Stafford Withington}
\affiliation{Cavendish Laboratory, University of Cambridge, JJ Thomson Ave,  Cambridge CB3 0HE, UK}

\author{John~J.~L.~Morton}
\affiliation{London Centre for Nanotechnology, UCL, 17-19 Gordon Street, London, WC1H 0AH, UK}
\affiliation{Department of Electrical and Electronic Engineering, UCL, Malet Place, London, WC1E 7JE, UK}

\maketitle

Solid state quantum devices typically operate at cryogenic temperates and find varied uses including as qubits~\cite{nakamura1999coherent,hendrickx2021four}, memories~\cite{kubo2011hybrid, zhong2017nanophotonic}, transducers~\cite{mirhosseini2020superconducting} and sensors~\cite{thiel2016quantitative}. A large proportion of these devices are composite structures and the different materials in these devices have different coefficients of thermal expansion (CTE). Without careful engineering, these devices are therefore strained at their operating temperature as the composite device contracts unevenly upon cooling. Straining solid state devices has disparate effects --- in semiconductors strain can lift valence band degeneracy reducing scattering and increasing carrier mobility~\cite{dobbie2012ultra}, it can be used to extend coherence times for defect spins~\cite{sohn2018controlling} and it broadens spin lines in hybrid devices~\cite{pla2018strain}. It is therefore desirable to be able to control strain, reducing it when undesirable, and designing strain profiles where it is a useful engineering tool. 

Impurity spins are widely used solid state quantum systems which are sensitive to strain. Shallow donors in silicon exhibit a hyperfine interaction~\cite{feher1959electron}, which we exploit to quantify changes in strain. Bismuth donors in silicon are described by the spin Hamiltonian;
\begin{equation}
\label{eq:ESR}
H/h= A\mathbf{I\cdot S}+\gamma_{\rm e}\mathbf{B_0\cdot S}+\gamma_{\rm Bi}\mathbf{B_0\cdot I},
\end{equation}
where $h$ is Planck's constant, $A=$~1.475~GHz~\cite{feher1959electron,morley2010initialization} is the (isotropic) hyperfine coupling constant, and $\mathbf{B_0}$ is the external magnetic field, while $\gamma_{\rm e} =  27.997(1)$~GHz/T and $\gamma_{\rm Bi} = 6.9(2)$~MHz/T  are respectively the gyromagnetic ratios of the bound electron and Bi nuclear spin~\cite{wolfowicz2013atomic}. The hydrostatic strain linearly changes the hyperfine constant~\cite{mansir2018linear} and therefore couples in to the Hamiltonian.

In this work we investigate the use of dielectric layers deposited by atomic layer deposition (ALD) on top of bismuth ion implanted silicon substrates as an approach to strain mitigation in prototypical microwave memories. 
ALD is chosen due to being a highly conformal deposition technique in principle capable of producing a uniform coating over the flat substrate. We use it to deposit nominally inert dielectric layers with the aim to modify only the strain environment. 
The different coefficients of thermal expansion of these materials mean that the nominally unstrained room temperature device becomes strained at cryogenic temperatures. 
Here we use ALD layers to reduce strain inhomogeneity in the top of the silicon, but with rational design, including local ALD deposition, this approach could in principle be generalised to allow for strain engineering. This would be compatible with other approaches to strain mitigation including control of electrode deposition to reduce cryogenic strain~~\cite{glowacka2014development,iosad1999optimization} or electrode deposition at cryogenic temperatures. 

\begin{figure}
    \centering
    \includegraphics[width=\linewidth]{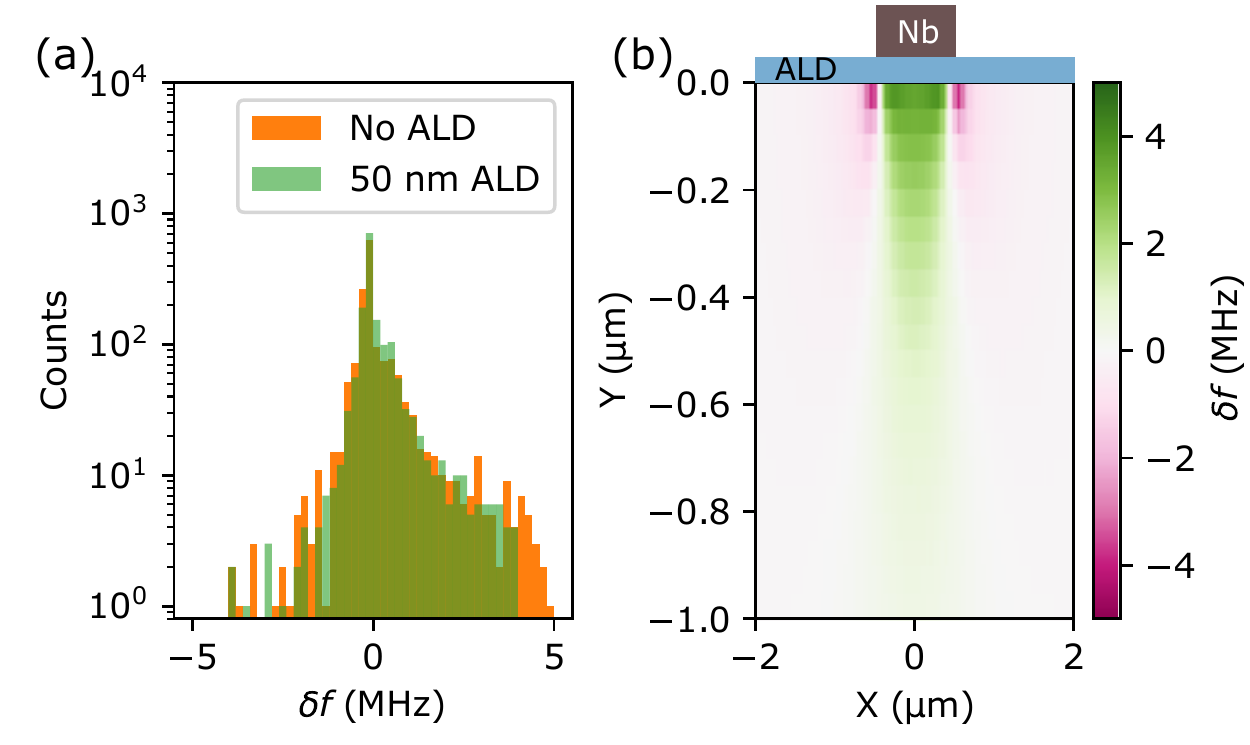}
    \caption{(a) Histogram showing simulated strain-induced frequency shifts in silicon underneath a 1~$\upmu$m wide, 100~nm thick Nb wire at 4K with and without a 50~nm ALD layer. (b) A map of the strain induced shifts which are binned to generate histograms in (a). The ALD and Nb wire are shown to scale on the top of the panel.}
    \label{fig:fig1}
\end{figure}

We perform finite element simulations of the cooling-induced strain in silicon samples coated by layers of Al$_2$O$_3$ with homogeneous Young's modulus and temperature dependent CTE given by crystalline Al$_2$O$_3$~\cite{yates1972thermal} (CTE of Al$_2$O$_3$ assumed temperature independent below 100~K due to lack of available data) using COMSOL. These simulations are presented in Fig.~\ref{fig:fig1} where we show the strain in the sample underneath a 1~$\upmu$m wide, 100~nm thick niobium lead in the presence and absence of a 50~nm thick ALD layer. We translate the strain into a shift in spin frequency assuming a maximal strain sensitivity (\dfdA~$= -5$). In Fig.~\ref{fig:fig1}a we bin the strain-induced shift for the silicon in the top 1~$\upmu$m of the silicon substrate and $\pm$2~$\upmu$m laterally from the centre of the niobium wire and show the resulting histogram. The spatially resolved strain map is shown in full in Fig.~\ref{fig:fig1}b. The addition of an ALD layer strains the surface, changing the mean strain whilst narrowing its distribution. Simulations like this motivated exploration of ALD layers for strain mitigation. 

The $I = 9/2$ nuclear spin of bismuth results in a splitting of 20 energy levels under application of an external static magnetic field. The energy levels are characterized by quantum numbers $F = I + S$ and its projection onto $\mathbf{B_0}$, $m_F$. There are two types of allowed transition between these levels, \sx~and \sz~which respectively have selection rules $\Delta F\Delta m_F = \pm1$ and $\Delta m_F = 0$. These correspond to $\mathbf{B_0}$ perpendicular and parallel to the diving microwave field. The transitions have varying sensitivity to magnetic field and hyperfine constant inhomogeneities which are characterized by \dfdb~and \dfdA~respectively. Some transitions undergo clock transitions in magnetic field~\cite{wolfowicz2013atomic, ranjan2020multimode} and hyperfine constant~\cite{mortemousque2014hyperfine}, where the transition frequency becomes insensitive to variations in these quantities to first order. Measuring the linewidth at these different clock transitions allows us to distinguish magnetic and strain broadening effects as broadening at a magnetic (hyperfine) clock transition is caused by strain (magnetic field) inhomogeneity. 

We fabricated composite devices where a silicon substrate is ion implanted by Bi$^+$ at a density of $\sim$10$^{17}$~cm$^{-3}$ in the top $\sim1~\upmu$m before being annealed for 5~mins at 900$^\circ$C~\cite{peach2018effect}. It is then coated by a 50~nm layer of Al$_2$O$_3$ deposited at 150$^\circ$C by ALD immediately after RCA cleaning. A 100~nm thick niobium resonator is then fabricated on top of the sample by lift off as in~\cite{o2020spin}. We compare this to a resonator with similar frequency fabricated without the ALD layer. We also perform the same measurement on the ALD sample when it is coated with a thick layer of photoresist ($\gtrsim2~\upmu$m). All resonators measured in this work have a coupling-limited quality factor of 5-20k due to the large coupling antenna optimal for the electron spin resonance measurements. 

Linewidths in these devices are determined by quasi-static inhomogeneities in the magnetic field and variations in strain. In silicon with natural isotopic abundance the $\sim$5\% $^{29}$Si causes $\sim$0.4~mT (11~MHz) of field broadening in bulk samples far from magnetic clock transitions~\cite{george2010electron}. In devices made on isotopically purified substrates the field broadening is low and linewidths are dominated by strain~\cite{pla2018strain}, and strain broadening is also observed in devices on natural silicon~\cite{george2010electron}. The spin transitions are measured at 100~mK by performing echo detected field sweeps (EDFS) using CPMG averaging using a home built electron spin resonance spectrometer as in Ref.~\cite{o2020spin}.

\begin{figure}
    \centering
    \includegraphics[width=\linewidth]{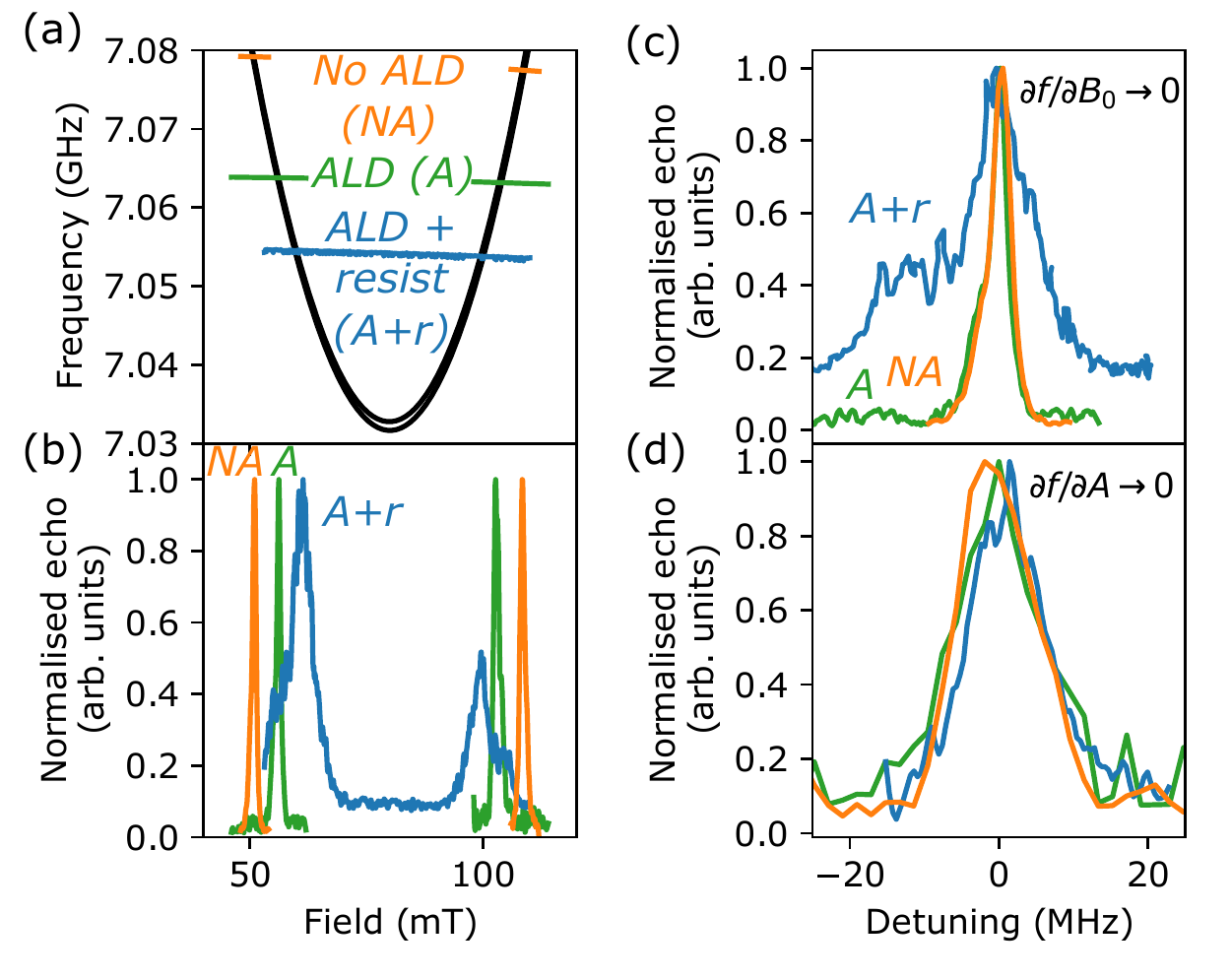}
    \caption{High power (+5dBm microwave power) EDFS of samples with and without ALD close to a magnetic field clock transition. (a) shows the resonator frequency relative to the transition frequency for the two resonators. (b) shows the echo amplitude as a function of field in these field sweeps. (c) shows the echo amplitude as a function of resonator/transition detuning showing that the ALD sample with photoresist has a substantially higher linewidth. (d) shows the echo amplitude as in (c) at the hyperfine clock transition, showing that here the resonators have the same linewidth.}
    \label{fig:fig2}
\end{figure}

We measure the spin linewidth at high drive power in all three device configurations in two limits: (i) close to a magnetic clock transition where \dfdb~$\sim -0.1\gamma_e$ and \dfdA$\sim -5$ (Fig.~\ref{fig:fig2} a-c) giving a strain-dominated lineshape and (ii) close to a hyperfine clock transition where \dfdb~$\sim\gamma_e$ and \dfdA~$\sim 0$ (Fig.~\ref{fig:fig2} d) giving a field dominated lineshape. 
At the magnetic clock transition the line shape of the resist-coated ALD sample is broad and shows a large shoulder at positive detuning from the centre of the spin line. This indicates that upon cooling the silicon has become inhomogeneously strained. Both the bare ALD coated sample and the no-ALD sample have much narrower linewidths showing that it is the resist causing this inhomogeneous strain, likely caused by thickness variations in the resist spun onto small pieces. 

At the hyperfine clock transition presented in Fig.~\ref{fig:fig2}d we show that all samples have a similar linewidth. This confirms our interpretation that the broad line in the resist coated sample is caused by strain --- if it were broadened by magnetic field inhomogeneity it would be much broader here due to the high sensitivity to magnetic fields. Instead we show that by correctly choosing the operating point we can eliminate the effects of strain present in our device. While it is much better to remove the strain by removing the resist causing the strain in these specific devices, it is a demonstration of how in general the hyperfine clock transition can be used to give line narrowing in composite devices. This may be useful for future applications where, for instance, magnetic field noise is removed by using isotopically purified samples and electric field noise and strain would otherwise become dominant. 

\begin{figure}
    \centering
    \includegraphics[width=\linewidth]{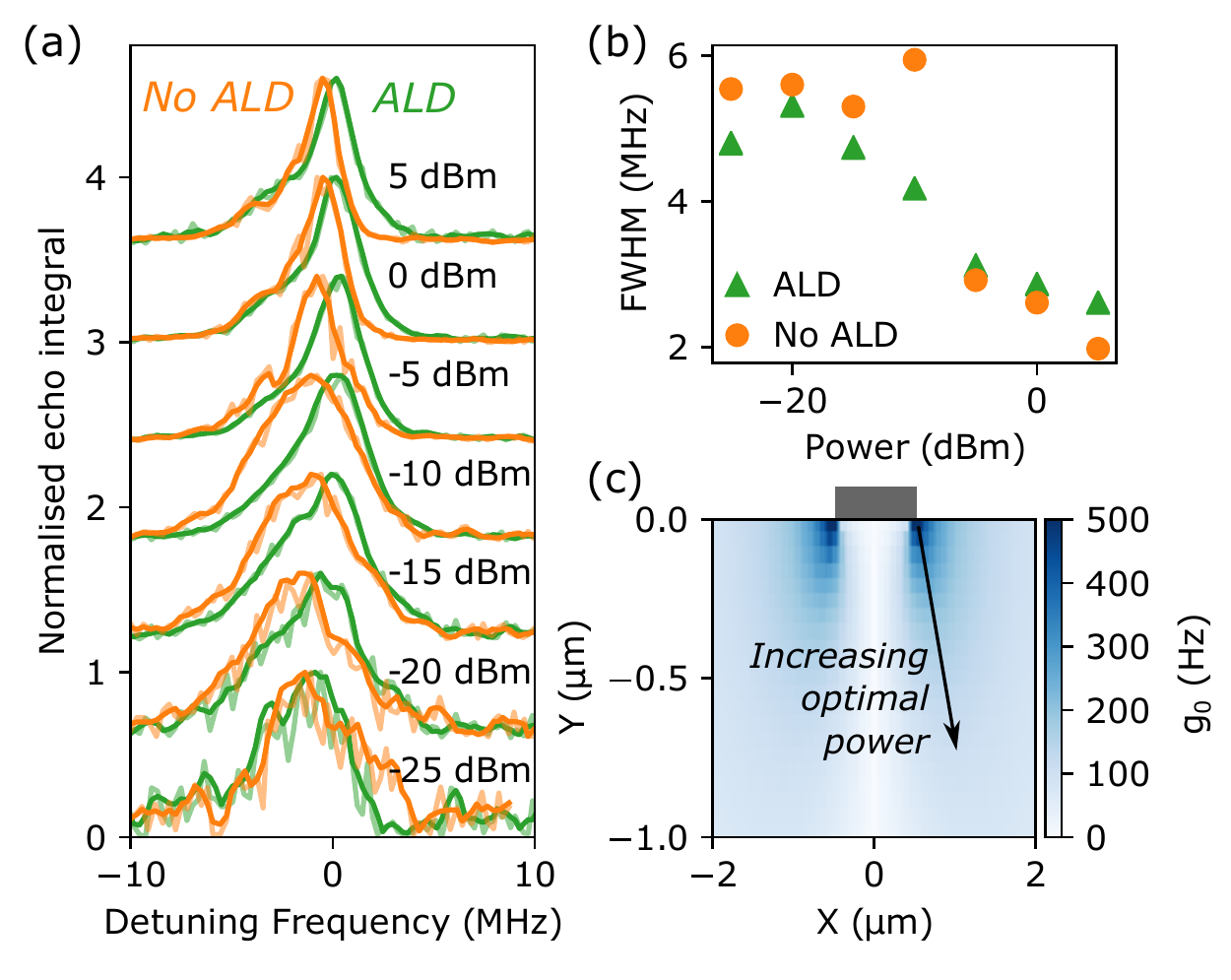}
    \caption{ (a) Power dependent EDFS of the ALD and no-ALD samples taken close to a magnetic field clock transition (\dfdb~$\sim$~0.1) plot as the difference in frequency between the simulated spin frequency and the measured resonator frequency. (b) The full width half max of the EDFS in (a) as a function of the measurement power. (c) Single spin coupling computed for a 100~$\Omega$ impedance resonator with a 1$~\upmu$m inductor (shown to scale). Regions with high (low) single spin coupling are more sampled at low (high) drive powers as they receive optimal tip angles. Spins with strong couplings are typically closer to the resonator and therefore more strained.  }
    \label{fig:fig3}
\end{figure}

In Fig.~\ref{fig:fig3}a we show power dependent EDFS and in Fig.~\ref{fig:fig3}b their full width half max (FWHM) of both the ALD and no ALD sample at virtually identical \dfdb~(-0.096$\gamma_e$ and -0.102$\gamma_e$ respectively). The similar and low \dfdb~means that lineshapes are determined by strain and their difference are caused by differences in the strain environment.
Performing EDFS at different microwave powers allows us to sample the spin linewidth for spins at different locations. We show a finite element calculation of the microwave field from a 100~$\Omega$ impedance resonator containing a 1~$\upmu$m inductor in  Fig.~\ref{fig:fig3}c. This shows how low (high) power measurements sample spins closer to (further from) the resonator which are more (less) strained due to their higher (lower) single spin coupling. 
In previous measurements of similar devices~\cite{o2020spin} close to an \sz~magnetic clock transition, we showed that at high drive powers the linewidths in natural silicon were dominated by nearest-neighbour silicon isotope mass shifts~\cite{sekiguchi2014host} which causes residual strain-independent broadening even at magnetic clock transitions of $\sim$2.5~MHz. 
In Ref.~\cite{o2020spin} we also showed that Rabi-like oscillations occurred in strained samples due to the correlation between strain-shifts and $B_1$ fields driving the spins.

The power dependent EDFS of the sample without ALD in Fig.~\ref{fig:fig3}a shows similar behaviour to this. At high drive powers it has a FWHM slightly above 2~MHz. There are Rabi-like oscillations in EDFS at negative detuning frequencies. At lower powers these oscillations disappear and instead the whole line broadens. The same high-power behaviour is seen for the ALD coated sample with a similar FWHM in Fig.~\ref{fig:fig3}b. Reducing the drive power has a different effect for the ALD sample - whilst the linewidth broadens, no oscillations appear in the EDFS and the linewidth broadens less than in the sample without the ALD coating. Indeed the FWHM for the ALD-coated sample is narrower than the sample without ALD at all powers lower than -5~dBm. 
The narrower lines at lower drive amplitudes imply that spins closest to the resonator are less strained in the sample with ALD than the sample without.
This qualitatively  agrees with the simulations presented in Fig.~\ref{fig:fig1} as ALD layers  and the finite element simulations show that for the sample without ALD we expect more spins with large positive shifts in frequency which is seen in the shoulder at positive frequency in the lowest power measurements of the sample without ALD. 
The contribution of nearest-neighbour-mass induced hyperfine shifts and residual magnetic-field broadening makes quantitative reproduction of the lineshapes (as in Refs.~\cite{ranjan2021spatially, pla2018strain}) challenging.
The lack of Rabi-like oscillations in the EDFS in the ALD sample means that the correlation between strain and $B_1$ (i.e. position relative to resonator wire) does not hold, which is not predicted by the finite element simulations shown in Fig.~\ref{fig:fig1}. This may be due to variations in the ALD layer along the length of the inductor wire and merits further investigation. 

As well as measuring static inhomogeneities from the spin-linewidth, we can measure fluctuating inhomogeneities (noise) which couples to the coherence times of the spins. In natural silicon the coherence time is typically limited by nuclear spin dynamics of $^{29}$Si which is modulated by \dfdb~and will limit our sensitivity to other noise. We measure the coherence times using WURST inversion pulses as high bandwidth inversion pulses which operate with high fidelity across spins coupled inhomogeneously to the resonator~\cite{sigillito2014fast}. 

\begin{figure}
    \centering
    \includegraphics[width=\linewidth]{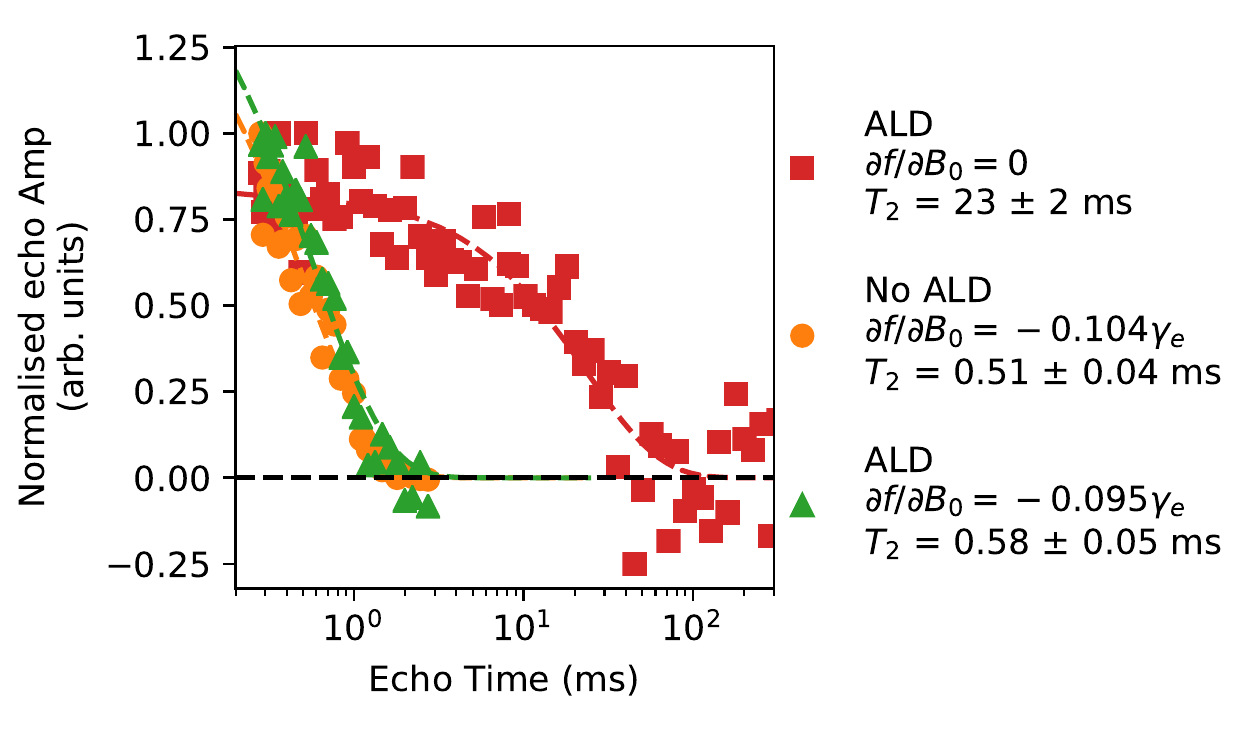}
    \caption{Comparison of the $T_2$ of the ALD-coated and no-ALD samples at similar fields (respectively 55.8~mT and 53.0~mT) and frequencies (respectively 7.064~GHz and 7.071~GHz) and consequently similar \dfdb. $T_2$ is very similar with discrepancies following the small change in \dfdb~implying that coherence times are limited by the same process --- the nuclei of $^{29}$Si as is commonly found for natural silicon. Also shown is $T_2$ of a different resonator on the ALD coated sample which precisely reaches the clock transition showing $T_2\sim$23~ms.}
    \label{fig:fig4}
\end{figure}

In Fig.~\ref{fig:fig4} we compare coherence-time measurements of the no-ALD sample and the ALD sample close to the magnetic clock transition (\dfdb~$\sim -0.1$) at points shown in Fig.~\ref{fig:fig2}. We show that the ALD-coated sample has coherence times slightly longer than the sample without ALD which we ascribe to differences in \dfdb. Both transitions show coherence times which are commensurate with Si:Bi samples with similar \dfdb~\cite{albanese2020radiative} where $T_2$ is limited by spectral diffusion from the nuclear spins of the $\sim 5\%~^{29}$Si. 
Using a different resonator with slightly lower frequency also on the ALD sample we bias the system precisely to a magnetic clock transition and measure the coherence time $T_2\sim23$~ms. This coherence time is very close to that predicted due to the bismuth-concentration dependent limit imposed by direct flip flops found in Ref.~\cite{wolfowicz2013atomic}. 
The ALD layer could in principle effect the coherence times by adding magnetic (electric) field noise which couple to the central spin with strength proportional to \dfdb~(\dfdA). 
If the ALD layer introduced substantial extra noise, then the spins in the ALD sample would show a decrease in $T_2$. We can therefore rule out significant noise caused by the ALD layer and conclude that induced magnetic field noise is inconsequential relative to $^{29}$Si nuclear spins and that electric field noise is sufficiently low to allow 23~ms coherence times at the clock transition. Future depth resolved studies may give more information about noise environments very close to the ALD layers. 

We explored two approaches to strain mitigation. One approach works in this system specifically where we can operate at strain-insensitive points. These points exist in some other systems but will depend on their Hamiltonians. The other approach is more general, depositing ALD layers to buffer the strain caused by narrow electrodes. These layers decrease strain in our composite devices particuarly close to the wires, although the measured reduction in linewidth remains modest.
This may be improved in future devices by selecting materials with higher Young's modulus. For instance the simulated broadening for diamond (as presented for Al$_2$O$_3$ in Fig.~\ref{fig:fig1}) has a narrower distribution of strain and maximal strain shift 1~MHz lower than for Al$_2$O$_3$. 
The lack of Rabi-like oscillations in EDFS from the ALD samples implies that strain does not depend just on proximity to the wire in these samples which may be caused by some imhomogeneity in the ALD layers which merits further investigation should this approach be used more widely.
Measuring the coherence time of spins in the ALD-coated samples close to magnetic clock transition allows us to conclude that magnetic or electric field noise from the ALD layer does not limit spin coherence. 
ALD layers alone are unlikely to be sufficient to completely remove strain in composite devices, however, they are a promising route to reduce strain and crucially are compatible with other methods of device engineering. For instance in the class of devices discussed in this paper, bismuth could be selectively implanted through a mask and resonators aligned to the implant. This would allow spins to be located in regions of quasi-homogeneous strain (e.g. underneath the resonator wire) and then the strain broadening within these regions to be further reduced by an ALD strain buffer. Furthermore, this approach may be generalised to allow strain profiles to be engineered as ALD layers can be selectively deposited by lift-off (e.g. in Ref.~\cite{casparis2018superconducting}).

\section{Acknowledgements}
The authors acknowledge useful discussions with Dr Mantas~\v{S}im\.{e}nas and Mr Joseph Alexander. The authors acknowledge the UK National Ion Beam Centre (UKNIBC) where the silicon samples were ion implanted and Dr Nianhua Peng who performed the ion implantation. This project has received funding from the U.K. Engineering and Physical Sciences Research Council, through UCLQ postdoctoral fellowships (O.~W.~K) Grant Number EP/P510270/1. JJLM acknowledges funding from the European Research Council under the European Union's Horizon 2020 research and innovation programme (Grant agreement No. 771493 (LOQO-MOTIONS).

\bibliography{bibliography}

\end{document}